\documentclass[12pt]{article}
\usepackage{amsmath}
\usepackage{amssymb}
\usepackage{amsfonts}
\usepackage{amscd}
\usepackage{delarray}
\usepackage{graphicx}

\usepackage{amsmath,amsmath,amsfonts,amssymb,color,bbm}

\usepackage{graphicx}

\def\Symp#1,#2,#3,#4.{\left[\left(\begin{array}{c}#1\\#2\end{array}\right),\left(\begin{array}{c}#3\\#4\end{array}\right)\right]}
\def\Vec#1,#2.{\left(\!\begin{array}{c}#1\\#2\end{array}\!\right)}
\def\vec#1,#2.{{#1\choose{#2}}}
\def\ket#1.{|#1\rangle}

\def\bra#1.{\langle#1|}
\def\braket#1,#2.{\langle#1|#2\rangle}
\newcommand{\beq}{\begin{equation}}
\newcommand{\eeq}{\end{equation}}
\newcommand{\beqa}{\begin{eqnarray}}
\newcommand{\eeqa}{\end{eqnarray}}

\begin{document}

\title{A wave-function model for the  $CP$-violation in mesons}
\date{}
\author{}
\maketitle
\vglue -1.8truecm
\centerline{M. Courbage\footnote
{\it  Laboratoire Mati\`ere et Syst\`emes
Complexes (MSC),  UMR 7057 CNRS et Universit\'e Paris 7- Denis
Diderot, Case 7056, B\^{a}timent Condorcet, 10, rue Alice Domon et
Léonie Duquet 75205 Paris Cedex 13, FRANCE.
\texttt{email: maurice.courbage@univ-paris-diderot.fr}}, T. Durt\footnote{TENA-TONA Vrije Universiteit Brussel, Pleinlaan 2, B-1050
Brussels, Belgium. \texttt{email:
thomdurt@vub.ac.be}} and S.M. Saberi Fathi\footnote{Laboratoire de Physique Th\'eorique et
Mod\`elisation (LPTM), UMR 8089, CNRS et Universit\'e de
Cergy-Pontoise, Site Saint-Matrin, 2, rue Adolphe Chauvin, 95302
Cergy-Pontoise Cedex, FRANCE. \texttt{email:
majid.saberi@u-cergy.fr}}
}

%



\begin{abstract}
In this paper we propose to associate a temporal two-component
wave-function to the decay process of meson particles. This simple
quantum model provides a good estimation of  the $CP$ symmetry
violation parameter. This result is based on our  previous paper
\cite{cds1} where we have  shown that the two-level Friedrichs
Hamiltonian model makes it possible to provide a qualitatively
correct phenomenological model of kaons physics. In this previous paper, we derived a violation parameter that is 14 times larger than the measured quantity. In the present paper we improve our estimation of the violation and obtain the right order of magnitude. The improvement
results from a renormalized superposition of the probability
amplitudes describing short and long exponential decays. The
renormalization occurs because the amplitudes that we are dealing
with are associated to the decay rate, and not to the integrated
decay rate or survival probability as is usually the case in
standard approaches to $CP$-violation. We also compare with recent
experimental data for the mesons $\mathrm{D}$ and $\mathrm{B}$ and
also there the agreement between our model and experimental data is
quite satisfying.
\end{abstract}

\maketitle

\section{Introduction}
The status of time in the quantum theory is still a controversial
subject \cite{booktime}. For instance, although standard quantum
mechanics allows us in principle to bring an unambiguous answer to
the question {\it is a certain particle located inside a given space
region at time t ?} (the answer is: the probability that the answer
is Yes is equal to the integral of the modulus square of the wave
function, at time t, over that region), it does not allow to answer
to the question {\it at what time will the particle enter this
region?} One could equivalently formulate this question in the form
{\it what is the probability that a particle hits a screen during a
given time interval?} Or {\it at what time will the particle hit the
screen?}, which justifies why the problem of deriving a temporal
distribution from Schr\"odinger's wave function (instead of a
spatial distribution) is often referred to in the literature as the
so-called {\it screen problem} \cite{mielnik,misrasud} or
arrival-time problem. One could believe that the source of the
confusion is that standard quantum mechanics is not a relativistic
theory and that in the framework of relativistic quantum field
theory these questions possess an unambiguous answer but this is not
the case. Even localization in space becomes problematic in quantum
field theory \cite{wignernewton}.

The screen problem is close to the tunneling phenomenon
\cite{cushing} which in turn presents many analogies with the decay
process \cite{arrival}; one could indeed modellize the decay of an
unstable state or resonance as a tunneling through the barrier of
potential that prevents instantaneous decay. Then, if we can predict
at which time the particle will tunnel, we can also predict the
distribution of decay times.

Although there is no unambiguous recipe for deriving temporal
probability distributions in the quantum theory, in the majority of
standard approaches to the decay process (among which the celebrated
Wigner-Weisskopf approach), metastable states or resonances are
formally characterized by a complex energy, of which the real part
is proportional to the mass of the particle and of which the
imaginary part is proportional to the inverse of the lifetime of the
resonance. For such exponential decay laws, the ``\emph{integrated
survival probability}" $P_s(t)$,  from an initial time 0 up to $t$,
obeys
\begin{equation}
P_s(t)=\frac{\psi^*(t)\psi(t)}{\psi^*(0)\psi(0)}.\label{exp}
\end{equation}
 Indeed, setting
\begin{equation}
\psi(t)=\psi(0) e^{-\mathrm{i}(m-\frac{\mathrm{i}}{2}\Gamma) t},
\end{equation}
we get \cite{gamow}, as a consequence of equation (\ref{exp}),

\begin{equation}
P_s(t)=e^{-{ t\over \tau}}\label{expdeux}
\end{equation}
 where the lifetime $\tau$  is  equal to the inverse of
the so-called decay rate $\Gamma$: $\Gamma={1\over \tau}$. These
exponential amplitudes are commonly used in particle physics.

It is well-known actually that this exponential decay-law is a mere
approximation and that one should expect discrepancies from it in
the short-time (Zeno regime) \cite{misrasud} and long-time behavior
of unstable quantum systems. Nevertheless, it is not easy to observe
experimentally such discrepancies \cite{zenoexp} because they mostly
affect non-significant parts of the statistical data that can be
collected about the distribution of decay-times and we shall not
consider these extreme cases in the rest of the paper.
Experimentally, it is most often easier to measure, instead of the
\emph{integrated survival probability}, the (temporal)
``\emph{density of probability}" of decay $p_d(t)$ or \emph{decay
rate} which is equal to minus the time derivative of $P_s(t)$:
$p_d(t)=d(1-P_s(t))/dt=-dP_s(t)/dt$.  In the case of an exponential decay
nevertheless this subtle point is often overlooked without
prejudicial consequences because $P_s(t)$ is {\it proportional} to
$p_d(t)$.

Indeed, deriving both sides of the expression (\ref{expdeux}) relatively to time, we obtain
\begin{equation}
p_d(t)=-\frac{dP_s(t)}{dt}=\Gamma\, e^{-\Gamma t}=p_d(0)\,
e^{-\Gamma t}.\label{exprate}
\end{equation}
One can also, still in the case of exponential decays, express the
decay rate in function of the norm of the state as follows
\begin{equation}
p_d(t)=-\frac{dP_s(t)}{dt}=p_d(0){\psi^*(t)\psi(t)\over
\psi^*(0)\psi(0)}.\label{exprategen}
\end{equation}

If one considers more complex behaviors, such as the superposition
of two exponential decay processes with different lifetimes,  then
$P_s(t)$ is \emph{no longer proportional} to $p_d(t)$! This is for
instance what occurs in experimental observations of $CP$-violation
during which populations of pairs of pions are estimated along the
decay  of unstable short lived and long lived kaons, and where an
interference effect is exhibited leading to $CP$-violation.

In the present paper we shall show that, if one makes a correct
superposition of the amplitude associated to the densities of the
short and long decay processes, the $CP$-violation parameter that we
derive  must be re-estimated. We  show how a simple model that we
developed in the past provides a correct prediction of the magnitude
of the observed $CP$-violation. We compare with more recent
experimental of the mesons $\mathrm{D}$ and $\mathrm{B}$.


\section{Phenomenology of kaon particles }

 Let us first recall some basic experimental facts. Kaons are bosons
that were discovered in the forties during the study of cosmic rays.
They are produced by collision processes in nuclear reactions during
which the strong interactions dominate. They appear in pairs
$\mathrm{K}^{0}$, $\overline{\mathrm{K}}^{0}$ \cite{perkins,hokim}.

The $\mathrm{K}$ mesons are eigenstates of the parity operator $P$:
$ P|\mathrm{K}^0\rangle=- |\mathrm{K}^0\rangle$, and $
P|\overline{\mathrm{K}}^0\rangle=- |\overline{\mathrm{K}}^0\rangle$.
$\mathrm{K}^0$ and $\overline{\mathrm{K}}^0$ are charge conjugate to
each other  $ C|\mathrm{K}^0\rangle=
|\overline{\mathrm{K}}^0\rangle$, and $
C|\overline{\mathrm{K}}^0\rangle= |\mathrm{K}^0\rangle$. We get
thus,
\begin{equation}
CP|\mathrm{K}^0\rangle= -|\overline{\mathrm{K}}^0\rangle, ~~~
 CP|\overline{\mathrm{K}}^0\rangle=
-|\mathrm{K}^0\rangle.
\end{equation}
Clearly $|\mathrm{K}^0\rangle$ and $|\overline{\mathrm{K}}^0\rangle$
are not $CP$-eigenstates, but the  following combinations
\begin{equation}\label{kk1}
\nonumber|\mathrm{K}_1\rangle=\frac{1}{\sqrt{2}}\big{(}|\mathrm{K}^0\rangle
-|\overline{\mathrm{K}}^0\rangle\big{)},~~~
|\mathrm{K}_2\rangle=\frac{1}{\sqrt{2}}\big{(}|\mathrm{K}^0\rangle
+|\overline{\mathrm{K}}^0\rangle\big{)},
\end{equation}
are $CP$-eigenstates:
\begin{equation}
CP|\mathrm{K}_1\rangle=+|\mathrm{K}_1\rangle, ~~~
CP|\mathrm{K}_2\rangle=-|\mathrm{K}_2\rangle.
\end{equation}
In the
absence of matter, kaons disintegrate through weak interactions
\cite{hokim}. Actually, $\mathrm{K}^0$ and $\overline{\mathrm{K}}^0$
are distinguished by their mode of \emph{production}. $\mathrm{K}_1$
and $\mathrm{K}_2$ are the decay modes of kaons. In absence of
$CP$-violation, the weak disintegration process distinguishes the
$\mathrm{K}_{1}$ states which decay only into ``$2\pi$" while the
$\mathrm{K}_{2}$ states decay into ``$3\pi, \pi e \nu, ...$"
\cite{leebook}. The lifetime of the $\mathrm{K}_{1}$ kaon is short
($\tau_{S}\approx 8.92\times10^{-11}~^\mathrm{s}$), while the
lifetime of the $\mathrm{K}_{2}$ kaon is quite longer
($\tau_{L}\approx 5.17\times10^{-8}~^\mathrm{s}$) \cite{perkins}.

$CP$-\emph{violation}  was discovered by Christenson et al.
\cite{christ}. $CP$-violation means that the long-lived kaon can
also decay to ``$2\pi"$. Then, the $CP$ symmetry is slightly
violated (by a factor of $10^{-3}$) by weak interactions so that the
$CP$ eigenstates $\mathrm{K}_1$ and $\mathrm{K}_2$ are not exact
eigenstates of the decay interaction. Those exact states are
characterized by lifetimes that are in a ratio of the order of
$10^{-3}$, so that they are called the short-lived state
($\mathrm{K}_S$) and long-lived state ($\mathrm{K}_L$ ). They can be
expressed as coherent superpositions of the $\mathrm{K}_1$ and
$\mathrm{K}_2$ eigenstates through
\begin{eqnarray}\label{kk2}
\nonumber|\mathrm{K}_L\rangle=\frac{1}{\sqrt{1+|\epsilon|^2}}\big{[}
\epsilon
~|\mathrm{K}_1\rangle + |\mathrm{K}_2\rangle \big{]},\\
|\mathrm{K}_S\rangle=\frac{1}{\sqrt{1+|\epsilon|^2}}\big{[}
|\mathrm{K}_1\rangle +\epsilon ~ |\mathrm{K}_2\rangle \big{]},
\end{eqnarray}
where $\epsilon$ is a complex $CP$-violation parameter,
$|\epsilon|\ll1$.
$\mathrm{K}_L$ and $\mathrm{K}_S$ are the eigenstates of the
Hamiltonian for the mass-decay matrix \cite{hokim,leebook} which has
 the following form in the basis $|\mathrm{K}^0\rangle$ and
$|\overline{\mathrm{K}}^0\rangle$ has the following form:
\begin{equation}
H=M-\frac{\mathrm{i}}{2}\Gamma\equiv
\begin{array}({cc}) M_{11}-\frac{\mathrm{i}}{2}\Gamma_{11} &
M_{12}-\frac{\mathrm{i}}{2}\Gamma_{12}  \\
M_{21}-\frac{\mathrm{i}}{2}\Gamma_{21} &
M_{22}-\frac{\mathrm{i}}{2}\Gamma_{22}
\end{array}
\end{equation}
where $M$ and $\Gamma$ are
individually Hermitian since they correspond to observables (mass
and lifetime). The corresponding eigenvalues of the mass-decay
matrix are equal to
\begin{equation}\label{mas}
E_L=m_L-\frac{\mathrm{i}}{2}\Gamma_L,~~~\mathrm{ and} ~~~
E_S=m_S-\frac{\mathrm{i}}{2}\Gamma_S.
\end{equation}
 We shall derive an explicit
expression of this mass-decay matrix in our model.

The $CP$-violation was established by the observation that
$\mathrm{K}_L$ decays not only via three-pion, which has natural
$CP$ parity, but also via the two-pion (``$2\pi$") mode with an
experimentally observed violation amplitude of the order of $10^{-3}$.

Let us now reconsider how the simple model (\ref{kk2}) and
(\ref{mas}) is related to the experimental data. A series of
detections is performed at various distances from the source of a
neutral kaon beam in order to estimate the variation of the
populations of emitted pion $\pi^+,\pi^-$ pairs in function of the
proper time. The initial state is a neutral kaon state
\begin{equation}\label{kaonstate}
|\mathrm{K}^0\rangle= \frac{1}{\sqrt{2}}\big{(}|\mathrm{K}_1\rangle
+|{\mathrm{K}}_2\rangle\big{)}=
\frac{\sqrt{1+|\epsilon|^2}}{\sqrt{2}(1+\epsilon)}\big{(}|\mathrm{K}_S\rangle
+|{\mathrm{K}}_L\rangle\big{)}.
\end{equation}
 As the $\pi^+,\pi^-$ pairs are $CP$-eigenstates
for the eigenvalue +1, their population is proportional to the
$\mathrm{K}_1$ population. In the case that $\epsilon$ equals 0 (no
$CP$-violation), $|\mathrm{K}_S\rangle=|\mathrm{K}_1\rangle$ and
$|\mathrm{K}_L\rangle=|\mathrm{K}_2\rangle$ so that, for times quite
longer than $\tau_S$, no $\pi^+,\pi^-$ pair is likely to be
observed. The experiment shows on the contrary that these pairs are
observed.


\subsection{Standard formulation of $CP$-violation}

The standard modeling of the process goes as follows: in accordance
with the expression (\ref{kaonstate}), the linearity of
Schr\"odinger equation, the transition amplitude $\psi_1(t)$ towards
the $\mathrm{K}_1$ state at time $t$ obeys
\begin{equation}
\psi_1(t)=\langle\mathrm{K}_1|\mathrm{K}^0(t)\rangle=\frac{\sqrt{1+|\epsilon|^2}}{\sqrt{2}
(1+\epsilon)}\left(\langle\mathrm{K}_1|\mathrm{K}_S(t)\rangle
+\langle\mathrm{K}_1|{\mathrm{K}}_L(t)\rangle\right)
\end{equation}
and by using
\begin{equation}
|\mathrm{K}_S(t)\rangle=|\mathrm{K}_S\rangle \,e^{-\mathrm{i}E_St},
~~~|\mathrm{K}_L(t)\rangle=|\mathrm{K}_L\rangle
\,e^{-\mathrm{i}E_Lt}
\end{equation}
we obtain
 \begin{equation}
\psi_1(t)=\frac{\sqrt{1+|\epsilon|^2}}{\sqrt{2}(1+\epsilon)}(\langle
\mathrm{K}_1|\mathrm{K}_S\rangle e^{-\mathrm{i}E_St}+ \langle
\mathrm{K}_1|\mathrm{K}_L\rangle
e^{-\mathrm{i}E_Lt}).\label{nonnorm}
\end{equation}
So that, combining with equation (\ref{kk2}) we get
\begin{equation}\label{psi}
\psi_1(t)=\frac{1}{\sqrt{2}(1+\epsilon)}\left( e^{-\mathrm{i} (m_S-\frac{\mathrm{i}}{2}\Gamma_S)
t}+ \epsilon e^{-\mathrm{i}
(m_L-\frac{\mathrm{i}}{2}\Gamma_L) t}\right)
\end{equation}
Now, some confusion exists in the literature regarding how to
proceed next. We tried to reconstitute a derivation of the generic
formula that we shall present here, having in mind that this
derivation presents certain weaknesses (that we shall address in a
separate publication), but, as we propose in a next section an
alternative approach to the problem, this does serve our purposes.

It is generally admitted \cite{perkins,hokim} that the intensity
$I(t)$ of $\pi^+,\pi^-$ pair detection at time $t$ obeys the
constraint
\begin{equation}
I(t)=|\psi_1(t)|^2.\label{exprategenkaon}
\end{equation}
Then we get by a straightforward computation
 \begin{eqnarray}\label{intens}
\nonumber I(t)={I(t=0)\over |1+\epsilon|^2}\,\bigg{(}e^{-\Gamma_S t}+
|\epsilon |^2e^{-\Gamma_L t}+
|\epsilon |e^{-({\Gamma_S +\Gamma_L\over 2})t} \cos
\big{(}\triangle m t+\arg(\epsilon)\big{)}\bigg{)}
\end{eqnarray}
where $I_0$ is a global factor constant in time, and $\triangle
m=|m_L-m_S|$,  so that an interference term is likely to appear,
that reveals the existence of a $CP$-violation. The existence of
such an effect was confirmed by experiments \cite{christ}. By
fitting this oscillating contribution with the observed data one
derives an estimation of the mass difference between the short and
long lived state as well as the phase of $\epsilon$ and its
amplitude. All this leads to an experimental estimation of
$\epsilon$ (that we shall denote $\epsilon^{\mathrm{exp}}$)
\cite{kk}
\begin{equation}\label{expepsilon}
|\epsilon^{\mathrm{exp}}|=(2.232\pm0.007)\times10^{-3}, ~~~
\mathrm{arg}(\epsilon^{\mathrm{exp}})=(43.5\pm 0.7)^\circ.
\end{equation}


\subsection{Alternative estimation of $|\epsilon|$\label{consist}}
There exist different experimental quantities that depend on the
$CP$ violation parameter. This means that there are alternative
experiments that make it possible to estimate the value of
$\epsilon^{\mathrm{exp}}$. An alternative way to estimate the
modulus of  $\epsilon^{\mathrm{exp}}$ consists of measuring ratios
of production rates. For instance, the ratio between the production
rate of charged pion pairs emitted by Long-lived states and the one
corresponding to Short-lived states obeys
\begin{equation}\label{ratio2}
{\mathrm{Proba.\, per\, unit\, of\, time }\,(K_L\rightarrow
\pi^+,\pi^-) \over \mathrm{Proba.\, per\, unit\, of\, time }\,(K_S
\rightarrow \pi^+,\pi^-)}={|\mathrm{Amplitude }\,(K_L\rightarrow
\pi^+,\pi^-)|^2 \over |\mathrm{Amplitude}\,(K_S\rightarrow
\pi^+,\pi^-)|^2}.
\end{equation}
Now,
\begin{eqnarray}\label{ratio}
\nonumber\epsilon &=&{\mathrm{Amplitude }\,(K_L\rightarrow K_1)
 \over \mathrm{Amplitude}\,(K_S\rightarrow K_1)}
 ={\mathrm{Amplitude }\,(K_L\rightarrow
K_1) \times\mathrm{Amplitude}\,(K_1\rightarrow \pi^+,\pi^-)
 \over \mathrm{Amplitude }\,(K_S\rightarrow
K_1) \times\mathrm{Amplitude}\,(K_1\rightarrow \pi^+,\pi^-)}\\
 &=&{\mathrm{Amplitude }\,(K_L\rightarrow
\pi^+,\pi^-)
 \over \mathrm{Amplitude}\,(K_S\rightarrow \pi^+,\pi^-)},
 \end{eqnarray}(where we made use of the fact that $ \pi^+,\pi^-$ are CP=+1 eigenstates like $K_1$ and belong to a space orthogonal to the CP=-1 eigenspace to which $K_2$ states belong),
 so that
\begin{equation}\label{ratio3}
{\mathrm{Proba.\, per\, unit\, of\, time }\,(K_L\rightarrow
\pi^+,\pi^-) \over \mathrm{Proba.\, per\, unit\, of\, time }\,(K_S
\rightarrow \pi^+,\pi^-)}=|\epsilon|^2
\end{equation}
The estimation of $|\epsilon|$ that is obtained by measuring this
ratio coincides with the value $|\epsilon^{\mathrm{exp}}|$ mentioned
above, which is also an indirect proof of the relevance and of the
consistency of the standard modeling of $CP$-violation.

\section{Theoretical model}

In 1957, Lee, Oehme and Yang (LOY) \cite{loy} derived a time
evolution equation of the neutral kaons using the Weisskopf-Wigner
approach to the decay of quantum systems \cite{ww}. LOY's Hamiltonian describes
$(\mathrm{K}^0,\overline{\mathrm{K}}^0 )$ evolution modes. Later on, the
LOY equation  has been improved by several authors
\cite{khalfin57,girad,khalfin, chsud,cds2} in order to obtain a
correction due to the departure from the pure exponential decay for
short and long times. Chiu and Sudarshan obtained a numerical estimate of the
modulus of the $CP$-violation parameter that is 30 times the
experimental value. Our approach, based on the derivation of a
master equation from a Friedrichs Hamiltonian \cite{fried,grecos} in
terms of  the decaying modes $(\mathrm{K}_1, \mathrm{K}_2)$ under
weak coupling approximation, provided a new estimate of the modulus
of the $CP$-violation parameter that is at first sight 14 times greater than the
experimental modulus while the estimated phase is roughly correct.

 The two-level Friedrichs interaction Hamiltonian couples two modes
and a continuous degree of freedom in such a way that the
Schr\"{o}dinger equation is \cite{fried,marchand,courbagem}
\begin{equation}
\left(\begin{array}{ccc} \omega_1 & 0& \lambda_1^*   \\ 0& \omega_2
& \lambda_2^* \\ \lambda_1 & \lambda_2 & \omega
\end{array}\right)\left(\begin{array}{ccc}f_1(t) \\ f_2(t) \\ g(\omega,t)
\end{array}\right)=\mathrm{i}\frac{\partial}{\partial t}
\left(\begin{array}{ccc}f_1(t) \\ f_2(t) \\ g(\omega,t)
\end{array}\right).\label{f1}
\end{equation}
In this model, the energies $\omega$ of the different modes of the
continuum range from $-\infty$ to $+\infty$. The masses
$\omega_{1(2)} $ represent the energies of the discrete levels, and
the factors $\lambda_{1(2)} $ represent the couplings to the
continuum of decay product. Consequently, the amplitudes of the
discrete and continuous modes obey
\begin{eqnarray}
& &\omega_1 f_1(t)+\lambda_1^*\int_{-\infty}^{\infty} d\omega
g(\omega,t)=\mathrm{i}\frac{\partial f_1(t)}{\partial t}
,~\label{f2}\\
& &\omega_2 f_2(t)+\lambda_2^*\int_{-\infty}^{\infty} d\omega
g(\omega,t)=\mathrm{i}\frac{\partial f_2(t)}{\partial
t},\label{f3}\\
& &\lambda_1 f_1(t)+\lambda_2 f_2(t)+\omega
g(\omega,t)=\mathrm{i}\frac{\partial g(\omega,t)}{\partial
t}.\label{f4}
\end{eqnarray}
Integrating the last equation  we obtain $g(\omega,t)$ assuming
$g(\omega,t=0)=0$ :
\begin{equation}
g(\omega,t)=-\mathrm{i} e^{-\mathrm{i}\omega t}\int_0^t d\tau
\big{[} \lambda_1 f_1(\tau)+\lambda_2 f_2(\tau)\big{]} v(\omega)
e^{\mathrm{i}\omega\tau},
\end{equation}
then, we substitute $g(\omega,t)$ in the above equation (\ref{f2})
we obtain
\begin{equation}
\mathrm{i}\frac{\partial f_1(t)}{\partial t}=\omega_1
f_1(t)-\mathrm{i}\lambda_1^*\int d\omega |v(\omega)|^2
e^{-\mathrm{i}\omega t}\int_0^t d\tau \big{[} \lambda_1
f_1(\tau)+\lambda_2 f_2(\tau)\big{]}
e^{\mathrm{i}\omega\tau},\label{f5}
\end{equation}
we also can obtain the same relation for $f_2(t)$ as
\begin{equation}
\mathrm{i}\frac{\partial f_2(t)}{\partial t}=\omega_2
f_2(t)-\mathrm{i}\lambda_2^*\int d\omega |v(\omega)|^2
e^{-\mathrm{i}\omega t}\int_0^t d\tau \big{[} \lambda_1
f_1(\tau)+\lambda_2 f_2(\tau)\big{]}
e^{\mathrm{i}\omega\tau}\label{f5-1}.
\end{equation}
One can  obtain \cite{cds1}
\begin{equation}
\mathrm{i}\frac{\partial }{\partial t} \left(\begin{array}{c}
f_1(t)\\f_2(t)
\end{array}\right)=
\left(\begin{array}{cc} \omega_1 - \mathrm{i} \pi |\lambda_1|^2 & -
\mathrm{i} \pi
\lambda_1\lambda_2^* \\
- \mathrm{i} \pi  \lambda_1^*\lambda_2& \omega_2 - \mathrm{i} \pi
|\lambda_2|^2
\end{array}\right)
\left(\begin{array}{c} f_1(t)\\f_2(t)
\end{array}\right).\label{fh1}
\end{equation}
Thus, we obtain an effective non-Hermitian Hamiltonian evolution,
$H_{\mathrm{eff}}=M-\mathrm{i}\frac{\Gamma}{2}$. The eigenvalues of
the above effective Hamiltonian  under the weak coupling constant
approximation become:
\begin{equation}
\omega_{+}=\omega_1-\mathrm{i}\pi|\lambda_1|^2+O(\lambda^4), ~~~
\omega_{-}=\omega_2-\mathrm{i}\pi|\lambda_2|^2+O(\lambda^4),
\end{equation}
In a first and very rough approximation, the eigenvectors of the
effective Hamiltonian are the same as the postulated kaons states.
\begin{equation}
| f_+\rangle=\begin{array}({c}) 1\\0
\end{array}=|\mathrm{K}_1\rangle ~~~\mathrm{and}~~~
| f_-\rangle=\begin{array}({c}) 0\\1
\end{array}=|\mathrm{K}_2\rangle.
\end{equation}

Phenomenology imposes that the complex Friedrichs energies $
\omega_{\pm}$ coincide with the observed complex energies. The
Friedrichs energies depend on the choice of the four parameters
$\omega_{1}$, $\omega_{2}$, $\lambda_{1}$ and $\lambda_{2}$ and the
observed complex energies are directly derived from the experimental
determination of four other parameters, the masses $m_{S}$ and
$m_{L}$ and the lifetimes $\tau_{S}$ and $\tau_{L}$. We must thus
adjust the theoretical parameters in order that they fit the
experimental data. This can be done by comparing the eigenvalue of
the effective matrix with the eigenvalue of the mass-decay matrix
which is taken in the equation (\ref{mas}). Finally, we have
\begin{eqnarray} \label{fp9}
\nonumber& &\omega_1=m_S ,~~~2\pi|\lambda_1|^2=\Gamma_S,\\
& & \omega_2=m_L, ~~~2\pi|\lambda_2|^2=\Gamma_L.
\end{eqnarray}
The above identities yields
\begin{equation}
\lambda_1=\sqrt{\frac{\Gamma_S}{2\pi}}\, e^{\mathrm{i}\theta_S},
~~~\lambda_2=\sqrt{\frac{\Gamma_L}{2\pi}}\, e^{\mathrm{i}\theta_L}
\end{equation}
where  $\theta_S$ and $\theta_L$ are real constants.


\subsection{$CPT$ invariance}
 Let us now discuss the $CPT$ invariance in our model. As
mentioned in the texts books like \cite{hokim,leebook}, $CPT$
invariance imposes some conditions on the mass-decay matrix, i.e.
\begin{equation}\label{cpt}
M_{11}=M_{22},~ \Gamma_{11}=\Gamma_{22},~ M_{12}=M^*_{21}~~\mathrm{
and}~~  \Gamma_{12}=\Gamma^*_{21}
\end{equation}
in the $\mathrm{K}^0$ and $\overline{\mathrm{K}}^0$ bases. But, we
note that our effective Hamiltonian is written in the $\mathrm{K}_1$
and $\mathrm{K}_2$ bases. Thus, we have to rewrite in the
$\mathrm{K}^0$ and $\overline{\mathrm{K}}^0$ bases. Thus, the
transformation matrix $T$ from the  $\mathrm{K}_1$ and $\mathrm{K}_2$
bases to the $\mathrm{K}^0$ and $\overline{\mathrm{K}}^0$ bases is
obtained  as
\begin{equation}
T=\frac{1}{\sqrt 2}\left(\begin{array}{cc} 1&1\\1&-1
\end{array}\right)=T^{-1}.
\end{equation}
Then, the effective Hamiltonian in the $\mathrm{K}^0$ and
$\overline{\mathrm{K}}^0$  bases,
$H_{\mathrm{eff}}^{0\overline{0}}$ is obtained by
\begin{equation}
H_{\mathrm{eff}}^{0\overline{0}}=TH_{\mathrm{eff}}T^{-1}=\frac{1}{2}\left(\begin{array}{cc}
1&1\\1&-1
\end{array}\right)
\left(\begin{array}{cc} \omega_1 - \mathrm{i} \pi |\lambda_1|^2 & -
\mathrm{i} \pi
\lambda_1\lambda_2^* \\
- \mathrm{i} \pi  \lambda_1^*\lambda_2& \omega_2 - \mathrm{i} \pi
|\lambda_2|^2
\end{array}\right)
\left(\begin{array}{cc} 1&1\\1&-1
\end{array}\right).\label{fh1-1}
\end{equation}
 we have, $H_{\mathrm{eff}}^{0\overline{0}} =$
\begin{equation}
\left(\begin{array}{cc} (m_S+ m_L) - \frac{\mathrm{i}}{2} \left(
\Gamma_S+\Gamma_L+2\sqrt{\Gamma_S\Gamma_L}\cos\triangle\theta\right),
&(m_S- m_L) - \frac{\mathrm{i}}{2} \left(
\Gamma_S-\Gamma_L+2\mathrm{i}\sqrt{\Gamma_S\Gamma_L}\sin\triangle\theta\right) \\
(m_S- m_L) - \frac{\mathrm{i}}{2} \left(
\Gamma_S-\Gamma_L-2\mathrm{i}\sqrt{\Gamma_S\Gamma_L}\sin\triangle\theta\right),
& (m_S+ m_L) - \frac{\mathrm{i}}{2} \left(
\Gamma_S+\Gamma_L-2\sqrt{\Gamma_S\Gamma_L}\cos\triangle\theta\right)
\end{array}\right)
.\label{fh1-2}
\end{equation}
where $\triangle\theta=\theta_S-\theta_L$. $CPT$ invariance
conditions in (\ref{cpt}) impose that
\begin{equation}
\triangle\theta=k\pi+\frac{\pi}{2},~~~ (k=\cdots,-1,0,1,\cdots).
\end{equation}
Here we choose $k=-1$, consequently,
$\triangle\theta=-\frac{\pi}{2}$. Then, we have
\begin{equation}
\begin{array}{ll}
M_{11}=M_{22}=(m_S+
m_L),&\Gamma_{11}=\Gamma_{22}=\Gamma_S+\Gamma_L,\\
M_{12}=M_{21}^*=(m_S-
m_L),&\Gamma_{12}=\Gamma_{21}^*=\Gamma_S-\Gamma_L-2\mathrm{i}\,\sqrt{\Gamma_S\Gamma_L}.
\end{array}
\end{equation}


\subsection{ $CP$-violation}
 Let us study in this
case the $CP$-violation. The Friedrichs model allows us to estimate
the value of $\epsilon$. For this purpose,  the effective
Hamiltonian (\ref{fh1}) acts on the $|\mathrm{K}_S\rangle$ vector
states (\ref{kk2}) as an eigenstate corresponding to the eigenvalue
$\omega_+=\omega_1-\mathrm{i}\pi\lambda_1^2=m_S-\mathrm{i}
\frac{\Gamma_S}{2}$, so that  we must impose that
$H_{\mathrm{eff}}\big{(}^{1}_{\epsilon}\big{)}=\omega_+ \big{
(}^{1}_{\epsilon}\big{)}$, from which we obtain after
straightforward calculations that
\begin{equation}
\epsilon=\frac{-\mathrm{i}\pi\lambda_1^*\lambda_2}{(\omega_2-\omega_1)
-\mathrm{i}\pi(|\lambda_2|^2-|\lambda_1|^2)}
\label{angle}\end{equation} and if we replace $\lambda$'s and
$\omega$'s by corresponding values in equation (\ref{fp9}) we have,
\begin{equation}
\epsilon=\frac{-\frac{\mathrm{i}}{2}\sqrt{\Gamma_L\Gamma_S}\,e^{\mathrm{i}\frac{\pi}{2}}}
{(m_L-m_S)-\frac{\mathrm{i}}{2}(\Gamma_L-\Gamma_S)}=\frac{\frac{1}{2}\sqrt{\Gamma_L\Gamma_S}}
{(m_L-m_S)-\frac{\mathrm{i}}{2}(\Gamma_L-\Gamma_S)}.\label{angle1}
\end{equation}
Similarly,  the effective Hamiltonian (\ref{fh1}) acts on the
$|\mathrm{K}_L\rangle$ vector states (\ref{kk2}) as an eigenstate
corresponding to the eigenvalue
$\omega_-=\omega_2-\mathrm{i}\pi\lambda_2^2=m_L-\mathrm{i}
\frac{\Gamma_L}{2}$, so that  we must impose that
$H_{\mathrm{eff}}\big{(}^{\epsilon}_{1}\big{)}=\omega_- \big{
(}^{\epsilon}_{1}\big{)}$, from which we obtain after
straightforward calculations that
\begin{equation}
\epsilon=\frac{\mathrm{i}\pi\lambda_1\lambda_2^*}{(\omega_2-\omega_1)
-\mathrm{i}\pi(|\lambda_2|^2-|\lambda_1|^2)}
\label{angle-1}\end{equation}
 and if we replace $\lambda$'s and
$\omega$'s by corresponding values in equation (\ref{fp9}) we have,
\begin{equation}
\epsilon=\frac{\frac{\mathrm{i}}{2}\sqrt{\Gamma_L\Gamma_S}\,e^{-\mathrm{i}\frac{\pi}{2}}}
{(m_L-m_S)-\frac{\mathrm{i}}{2}(\Gamma_L-\Gamma_S)}=\frac{\frac{1}{2}\sqrt{\Gamma_L\Gamma_S}}
{(m_L-m_S)-\frac{\mathrm{i}}{2}(\Gamma_L-\Gamma_S)}.\label{angle1-1}
\end{equation}
$\mathrm{K}_S$ and $\mathrm{K}_L$
provide the same expression for $\epsilon$ as can be seen from the equations
(\ref{angle1}) and (\ref{angle1-1}).

At this level we introduce a fundamentally new ingredient that constitutes
a breakthrough relatively to standard text-book approaches. This
ingredient consists of associating a temporal two-component
wave-function to the decay rate. It is based on an analogy with spin
1/2 spatial wave-functions that we present now.


\subsection{The spin 1/2 analogy}
Let us consider the two-components of the Pauli
wave function $(\Psi_1(x,T),\Psi_2(x,T))$ associated to a spinor at time $T$. The probability to find spin ``up" ((1,0)) (spin ``down" ((1,0)) ) at time $T$ in the
interval $[x,x+dx]$ is, according to the usual rules of Quantum
Mechanics, equal to $dx$ times
$|\Psi_1(x,T)|^2$ ($|\Psi_2(x,T)|^2$).

Let us now replace space by time, the spin operator by the $CP$
pseudo-spin operator, and the measurement of the position of a
particle by the measurement of the time at which decay occurs. The modulus square of the first
component of the wave function is then equal to the decay rate in
the $CP=+1$ sector. An initial $\mathrm{K}_0$ state is seen to
correspond, in virtue of this analogy, to a fifty-fifty coherent
superposition state of a Short state ${1\over
\sqrt{1+|\epsilon|^2}}\left(^1_\epsilon\right)$ and of a Long state
${1\over \sqrt{1+|\epsilon|^2}}\left(^\epsilon_1\right)$. The $CP=+1$ and
$-1$ components of the temporal wave function
$(\widetilde{\psi}_1(t),\widetilde{\psi}_2(t))$ associated to this
state are thus equal to
\begin{equation}
\widetilde{\psi}_1(t)={1\over \sqrt{2}\widetilde{N}}
\left(\sqrt{\Gamma_S}e^{-\mathrm{i}(m_S-\frac{\mathrm{i}}{2}\Gamma_S)
t} +\epsilon \sqrt{\Gamma_L}
e^{-\mathrm{i}(m_L-\frac{\mathrm{i}}{2}\Gamma_L) t}\right)
\end{equation}
and
\begin{equation}
\widetilde{\psi}_2(t)={1\over \sqrt{2}\widetilde{N}} \left( \epsilon
\sqrt{\Gamma_S}e^{-\mathrm{i}(m_S-\frac{\mathrm{i}}{2}\Gamma_S) t}
+\sqrt{\Gamma_L}e^{-\mathrm{i}(m_L-\frac{\mathrm{i}}{2}\Gamma_L )
t}\right).
\end{equation}

The probability to find a $CP=+1$ kaon (or $\mathrm{K}_1$ particle) in
the temporal interval $[t,t+dt]$ is, according to the usual rules of
quantum mechanics, equal to:
\begin{equation}
|\widetilde{\psi}_1(t)|^2={1\over
2\widetilde{N}^2}\,\left(\Gamma_Se^{-\Gamma_St}+|\epsilon|^2\Gamma_L
e^{-\Gamma_Lt} +2Re\left(\epsilon \sqrt{\Gamma_S\,
\Gamma_L}e^{-(i(m_L-m_S)+{\Gamma_S+\Gamma_L\over
2})t}\right)\right)\,dt.
\end{equation}
Similarly, the probability to find a $CP=-1$ kaon (or $\mathrm{K}_2$
particle) in the interval $[t,t+dt]$ is equal to:
\begin{equation}
|\widetilde{\psi}_2(t)|^2={1\over
2\widetilde{N}^2}\,\left(|\epsilon|^2\Gamma_Se^{-\Gamma_St}+\Gamma_Le^{-\Gamma_Lt}
+2Re\left(\epsilon \sqrt{\Gamma_S\,
\Gamma_L}e^{-(i(m_S-m_L)+{\Gamma_S+\Gamma_L\over
2})t}\right)\right)\,dt.
\end{equation}
Normalization is imposed by the requirement that (i) at time 0, when
a particle is prepared, its survival probability is equal to 1, (ii)
the survival probability tends to 0 when time tends to infinity,
which means that $\int_0^{\infty}dt (-)dP_s(t)/dt=\int_0^{\infty}dt
p_d(t)=1$, and (iii) $p_d(t)=\psi^*(t)\psi(t)$, in analogy with the
normalization condition that is imposed to spatially extended wave
functions in first quantization procedure.  $\widetilde N$ is thus
chosen in order to normalize the probability of decay to 1:
$\int_0^\infty\left(|\widetilde{\psi}_1(t)|^2+|\widetilde{\psi}_2(t)|^2\right)dt=1$.
This means that
\begin{equation}
 \widetilde{N}^2=1+|\epsilon|^2+\left({\sqrt{\Gamma_S\,
\Gamma_L}\,(\Gamma_S+\Gamma_L)\over (\triangle
m)^2+({\Gamma_S+\Gamma_L\over 2})^2}\right)\,Re(\epsilon).
\label{appen}\end{equation}


\subsection{Renormalization of the violation parameter}
Let us now reconsider $CP$-violation, having in mind the
wave-function model described in the previous section. In the case
that Short and Long decay processes coherently interfere, our
normalization criterion imposes that the {\emph effective}
transition amplitude $\widetilde{\psi}_1(t)$ towards the
$\mathrm{K}_1$ state at time $t$,  is equal to
\begin{eqnarray}\label{aaa1}
\nonumber\widetilde{\psi}_1 &= &\frac{1}{\widetilde N}\left({\langle
\mathrm{K}_1|\mathrm{K}_S\rangle\, \sqrt {-2\mathrm{Im}(E_S)}\,\,e
^{-\mathrm{i}E_St}+\langle \mathrm{K}_1|\mathrm{K}_L\rangle\,
\sqrt{ -2\mathrm{Im}(E_L)}}\,\,  e^{-\mathrm{i}E_Lt}\right)\\
& =&\frac{1}{\widetilde N}\left( \sqrt {\Gamma_S}\,\,e
^{-\mathrm{i}E_St}+\epsilon\, \sqrt{\Gamma_L}\,\,
e^{-\mathrm{i}E_Lt}\right)
\end{eqnarray}
where $\widetilde{N}$ obeys (\ref{appen}). Because of our new choice of
normalization, there appear new normalization factors
$\sqrt{-2Im(E_{L(S)})}$ that were not present at the level of
equation (\ref{nonnorm}). Thus, the theoretically estimated
intensity $I(t)$ has now the following form:
\begin{eqnarray}\label{in1}
\nonumber I(t) &=&{I(t=0)\over |1+\epsilon^{\mathrm{th}}|^2}\bigg(
\,e^{-2\pi \lambda_1^2t}+ |\epsilon^{\mathrm{th}}|^2\, e^{-2\pi
\lambda_2^2t}+ 2 |\epsilon^{\mathrm{th}}|\, e^{-\pi(
\lambda_1^2+\lambda_2^2)t}\cos(\triangle\omega
t+\arg(\epsilon^{\mathrm{th}}))\bigg)\\
 &=&{I(t=0)\over
|1+\epsilon^{\mathrm{th}}|^2}\,\bigg{(}e^{-\Gamma_S t}+
|\epsilon^{\mathrm{th}} |^2e^{-\Gamma_L t}+ |\epsilon^{\mathrm{th}}
|e^{-({\Gamma_S +\Gamma_L\over 2})t} \cos \big{(}\triangle m
t+\arg(\epsilon)\big{)}\bigg{)}
\end{eqnarray}
where we
define  $\epsilon^{\mathrm{th}}$ by the renormalisation condition

\begin{equation}\label{renorm}\epsilon^{\mathrm{th}}=\epsilon\, \sqrt{\frac{ \Gamma_L}{\Gamma_S}},\end{equation} so that
\begin{equation}
\epsilon^{\mathrm{th}}=\epsilon\,\,\sqrt{\frac{\Gamma_L}{\Gamma_S}}=\frac{\Gamma_{L}}{\Gamma_{S}}\,
\frac{\frac{1}{2}}{\frac{\triangle
m}{\Gamma_S}-\mathrm{i}\frac{\triangle\Gamma}{2\Gamma_S}}.\label{re1}
\end{equation}

\section{Experimental confirmations}
\subsection{Kaons}

By using the experimental ratio
$\frac{(m_L-m_S)}{-(\Gamma_L-\Gamma_S)}\approx \triangle m
\tau_S\approx 0.5$ and the above experimental values of
$\Gamma_L,\Gamma_S, m_L$, $m_S$, we obtain the following estimated
value for $\epsilon^{\mathrm{th}}$:
\begin{equation}\label{re11}
\epsilon^{\mathrm{th}}=\left({1.82\over \sqrt 2}\times 10^{-3}\right)\,\times\,
e^{\mathrm{i}(46.77)^\circ}\approx 0.6~\epsilon^{\mathrm{exp}}
\end{equation}
which shows that our simple model gives a rather good rough
estimation of $CP$-violation.

Actually, the Friedrichs model still possesses adjustable quantities
like the  cut-off of the coupling constants $\lambda_{1,2}$.
Different choices for the cut-off lead to slightly different
estimations of $\epsilon$ as we have shown in Ref.\cite{cds2} so
that at this level we are fully satisfied if we obtain a rough
agreement between our predictions for $\epsilon^{\mathrm{th}}$ and
its experimental counterpart $\epsilon^{\mathrm{exp}}$.


\subsection{Internal consistency of the renormalization prescription for $\epsilon$}
As we mentioned in the section \ref{consist}, there exist different
experimental quantities that depend on $\epsilon$. In order to
establish the self-consistency of our approach it is important to
check that these quantities are renormalized in a similar fashion.

Let us check that it is well so, repeating the resoning of the
section \ref{consist} in the wave-function approach. In the
wave-function approach we find that the amplitude that
$\mathrm{K}_L$ decays in the $CP=+1$ sector is weighted by a factor
$\sqrt{\Gamma_L}$. Similarly, the amplitude that $\mathrm{K}_S$
decays in the $CP=+1$ sector must be renormalized by a factor
$\sqrt{\Gamma_S}$.
\begin{equation}
 \frac{\mathrm{Production\, rate\, of}\,(\pi^+,\pi^-)\,
\mathrm{from}\,\mathrm{K}_L}{ \mathrm{Production \,rate \,
of}\,(\pi^+,\pi^-)\, \mathrm{from}\,\mathrm{K}_S}=\frac{\mathrm{Proba.\, per\, unit\, of\, time }\,(\mathrm{K}_L\rightarrow
\pi^+,\pi^-)}{\mathrm{Proba.\, per\, unit\, of\, time}\,(\mathrm{K}_S\rightarrow
\pi^+,\pi^-)}=\frac {|\epsilon|^2
\Gamma_L}{\Gamma_S} =|\epsilon^{\mathrm{th}}|^2.
\end{equation}
As we see from the previous equation, in the wave-function approach,
the same renormalization condition (\ref{renorm}) is consistently
used that we measure $CP$-violation through interference effects or
through ratios of production rates. This establishes the consistency
of our approach.

Our model also possesses some predicting power for what concerns
other particles like B and D particles, because the $CP$-violation
parameter is related to other quantities (life times and masses)
through the constraint (\ref{angle1}) as we shall now show.


\subsection{B-mesons}
The other example  is the  $CP$-violation in
the decay of $\mathrm{B}^0_s$  and $\overline{\mathrm{B}}^0_s$. The
experimental values are \cite{amslerB}
\begin{equation}\label{bb1}
\frac{\triangle\Gamma_s}{2\Gamma_s}=0.069^{+0.058}_{-0.062},~~~~\frac{1}{\Gamma_s}=1.470^{+0.026}_{-0.027}\,\,\mathrm{ps},
\end{equation}
or equivalently ($\Gamma_{L,H}=\Gamma_s\pm\triangle\Gamma_s/2$),
\begin{equation}\label{bb2}
\frac{1}{\Gamma_L}=1.419^{+0.039}_{-0.038}\,\,\mathrm{ps},~~~~\frac{1}{\Gamma_H}=1.525^{+0.062}_{-0.063}\,\,\mathrm{ps},
\end{equation}
and the difference of masses is
\begin{equation}\label{bb3}
\triangle m=17.7^{+6.4}_{-2.1}\,\,\mathrm{ps}^{-1}
\end{equation}
and the experimental $CP$-violation parameter of the $\mathrm{B}$
meson is \cite{amslerB,amslerCP}:
\begin{equation}\label{bb4}
\mathcal{A}_{SL}^{\mathrm{exp}}\simeq4\mathcal{R}e(\epsilon^{\mathrm{exp}}_B)=(-0.4\pm5.6)\times10^{-3}\Rightarrow
\left|\frac{q}{p}\right|^{\mathrm{exp}}=1.0002\pm0.0028.
\end{equation}
where
$\frac{\mathcal{A}_{SL}^{\mathrm{exp}}}{2}\approx1-\left|\frac{q}{p}\right|^{\mathrm{exp}}$.
By replacing in the equation (\ref{re1}) we obtain:
\begin{equation}\label{bb5}
\epsilon^{\mathrm{th}}_B=\frac{\Gamma_{L}}{\Gamma_{H}}\,
\frac{\frac{1}{2}}{\frac{\triangle
m}{\Gamma_s}-\mathrm{i}\frac{\triangle\Gamma_s}{2\Gamma_s}}=0.018 +
0.047\times10^{-3} \, \mathrm{i}
\end{equation}
Thus, our theoretical $\left|\frac{q}{p}\right|^{\mathrm{th}}$
prediction is:
\begin{equation}\label{bb6}
\left|\frac{q}{p}\right|^{\mathrm{th}}=\left|\frac{1-\epsilon^{\mathrm{th}}}{1+\epsilon^{\mathrm{th}}}\right|=0.96
\end{equation}
which is in fairly good agreement with the experimental value.


\subsection{D-mesons}
The other example is the $CP$-violation in
the decay of $\mathrm{D}$ meson. The experimental values for
$CP$-violation of
$\mathrm{D}^0\rightarrow\mathrm{K}_S^0\,\pi^+\,\pi^-$ as reported by
Belle \cite{amslerD} are as follows:
\begin{eqnarray}\label{dd1}
\frac{\triangle\Gamma}{2\Gamma}=\left(0.37\pm0.25^{+0.07+0.07}_{-0.13-0.08}\right),\\
\frac{\triangle
m}{\Gamma}=\left(0.81\pm0.30^{+0.10+0.09}_{-0.07-0.16}\right)
\end{eqnarray}
where  $1/\Gamma=\tau, ~(\hbar=1)$ is the mean life time
\begin{equation}\label{dd2}
\frac{1}{\Gamma}=\tau=\frac{\tau_{\overline{\mathrm{D}}^0}+\tau_{\mathrm{D}^0}}{2}=(410.1\pm1.5)\times
10^{-3}\,\mathrm{ps}
\end{equation}
The  $CP$-violation parameters are experimentally denoted by
$\left(\frac{q}{p}\right)$ and given by:
\begin{equation}\label{cp111}
\left|\frac{q}{p}\right|^{\mathrm{exp}}=\left|\frac{1-\epsilon^{\mathrm{exp}}}{1+\epsilon^{\mathrm{exp}}}\right|
=\left(0.86^{+0.30+0.06}_{-0.29-0.03}\right)
\end{equation}
 and
 \begin{equation}\label{cp112}
\phi^{\mathrm{exp}}=\arg\left(\frac{q}{p}\right)^{\mathrm{exp}}
=\arg\left(\frac{1-\epsilon^{\mathrm{exp}}}{1+\epsilon^{\mathrm{exp}}}\right)
=\left(-14^{+16+5+2}_{-18-3-4}\right)^\circ.
\end{equation}
 By replacing in the expression (\ref{re1}) we obtain
 \begin{equation}
\epsilon^{\mathrm{th}}=\left(0.077+ 0.035 \mathrm{i}\right).
 \end{equation}
Consequently,
 \begin{equation}
\left|\frac{q}{p}\right|^{\mathrm{th}}=0.86,~~~\phi^{\mathrm{th}}=-4.02^\circ.
\end{equation}
which is in fairly good agreement with the experimental value.


\section{Conclusions} \label{Conclusions.}
As we have shown, if one takes fully account of the subtle
distinction between decay rate and integrated survival probability, the $CP$-violation parameter that we derive from the
experiment must be re-estimated, on the basis of our spatial-temporal wave function analogy. We showed how a simple model that
we developed in the past \cite{cds1,cds2} provides a correct
prediction of the magnitude of the observed $CP$-violation. We also
applied this model to other particle decay data that also reveal
$CP$-violation and discussed the accordance between theoretical
predictions and observations in those cases.

It is worth noting that the problem of associating a temporal
distribution to a superposition of exponential decay processes is
intimately related to the possibility of defining a Time Operator in
quantum mechanics, which is also a controversial question. Pauli
showed thanks to very simple arguments that if one could find an
operator $\hat{T}$ that satisfies canonical commutation rules
$[\hat{T}, \hat{H}]=i\hbar$ with the Hamiltonian operator $\hat{H}$
of a quantum system, then the spectrum of $\hat{H}$ ought to be
unbounded by below, which clearly constitutes a physical
impossibility. A possible way to answer Pauli's objection is to
define a ``super'' time operator that acts onto density matrices
rather than onto pure states. It is not our purpose to investigate
this question in the present work, although the results that we
derive here were to a large extent inspired by our study of the time
operator (this work will be presented in a separate publication).
The present approach however does not presuppose the existence of a
Time Operator. Neither does it rely on a particular interpretation
of the Quantum Theory\cite{cs,cour80,MPC}. In that separate publication (still in preparation), we show
that, in the Wigner-Weisskopf approximation, when the energy
spectrum extends from $-\infty$ to $+ \infty$, the exponential
amplitude probability  is obtained from a time operator
representation of the wave function, where the Time (super)operator
is defined as conjugated to the Hamiltonian (super)operator. It is a
Fourier transform of a resonance with a complex pole in the energy
representation. The idea that underlies the (super)time operator
 approach is that Pauli's objections are valid in the Hilbert space of
pure states, but are not valid in the (super)operator space.


\medskip                      
   \leftline{\large \bf Acknowledgment}
\medskip T.D. acknowledges support from the
ICT Impulse Program of the Brussels Capital Region (Project Cryptasc), the IUAP programme of the Belgian government, the grant V-18, and the Solvay Institutes for
 Physics and Chemistry. Thanks to Jean-Marie Frere (ULB) and Pascal David (Paris 7) for helpful comments and precious informations.

\end{document}